\documentclass[11pt,english,aps,pra,onecolumn,tightenlines,groupedaddress,notitlepage,floatfix,fleqn]{revtex4-1}
\pdfoutput=1

\usepackage[T1]{fontenc}
\usepackage{amsmath}
\usepackage{amssymb}
\usepackage{amsfonts}
\usepackage{bbm}
\usepackage{epsfig}
\usepackage{grffile}

\usepackage[usenames,dvipsnames]{color}
\definecolor{dblue}{rgb}{0,0.1,.6}

\usepackage[colorlinks=true,citecolor=dblue,linkcolor=dblue,urlcolor=dblue]{hyperref}
\usepackage[all]{hypcap}
\newcommand{\Footnote}[1]{\footnote{\unexpanded{#1}}}

\usepackage{setspace}

\newcommand{\id}{\mathbbm{1}}
\newcommand{\Tr}{\operatorname{Tr}}
\newcommand{\dist}{\operatorname{dist}}
\newcommand{\bra}{\langle}
\newcommand{\ket}{\rangle}
\newcommand{\mc}[1]{\mathcal{#1}}

\renewcommand{\vec}[1]{{\boldsymbol{#1}}}
\newcommand{\norm} [1]{\left\Vert #1\right\Vert}

\newcommand{\G} {\mathcal{G}}
\newcommand{\bG}{\bar{\mathcal{G}}}
\newcommand{\vs}{\vec{\sigma}}
\newcommand{\bs}{\bar{\sigma}}
\newcommand{\trunc}{\text{trunc}}
\newcommand{\SVD}{\text{SVD}}
\newcommand{\SVT}{\text{SVT}}
\newcommand{\diagt}{\text{diagt}}

\usepackage{titlesec}
\titleformat{\section}[block]
  {\fontsize{12}{15}\bfseries}
  {\thesection.}
  {1em}
  {\MakeUppercase}
\titleformat{\subsection}[hang]
  {\fontsize{12}{15}\bfseries}
  {\thesection~\thesubsection.}
  {1em}
  {}

\newcommand{\duke} {Department of Physics, Duke University, Durham, NC 27708, USA}

\begin{document}

\title{The matrix product approximation for the dynamic cavity method}
\author{Thomas Barthel\vspace{-0.8em}}
\affiliation{\duke}

\date{November 11, 2019}

\begin{abstract}
Stochastic dynamics of classical degrees of freedom, defined on vertices of locally tree-like graphs, can be studied in the framework of the dynamic cavity method which is exact for tree graphs. Such models correspond for example to spin-glass systems, Boolean networks, neural networks, and other technical, biological, and social networks. The central objects in the cavity method are edge messages -- conditional probabilities of two vertex variable trajectories.
In this paper, we discuss a rather pedagogical derivation for the dynamic cavity method, give a detailed account of the novel matrix product edge message (MPEM) algorithm for the solution of the dynamic cavity equation as introduced in \emph{Phys.\ Rev.\ E} \textbf{97}, 010104(R) (2018), and present optimizations and extensions.
Matrix product approximations of the edge messages are constructed recursively in an iteration over time. Computation costs and precision can be tuned by controlling the matrix dimensions of the MPEM in truncations. Without truncations, the dynamics is exact. Data for Glauber-Ising dynamics shows a linear growth of computation costs in time. In contrast to Monte Carlo simulations, the approach has a much better error scaling. Hence, it gives for example access to low probability events and decaying observables like temporal correlations. We discuss optimized truncation schemes and an extension that allows to capture models which have a continuum time limit.
\end{abstract}

\maketitle
\vspace{-1.3em}
\begin{spacing}{0.88}
\tableofcontents
\end{spacing}

\section{Introduction}
Statistical physics provides a comprehensive framework for the study of many-body systems in equilibrium with their environment. It is a foundation of modern physics, tracing back the laws of thermodynamics to characteristics of the microscopic degrees of freedom. Beyond physics, its formalism is successfully applied in many different fields such as information theory, computer science, economy, and sociology. In contrast, our methodological toolbox for dynamics in probabilistic systems is rather limited, which restricts our understanding of stochastic dynamical phenomena.

Stochastic dynamics, i.e., dynamics that are governed by probabilistic instead of deterministic rules, are ubiquitous in nature as well as in social and technological systems \cite{VanKampen2007,Freund2010,Capasso2004,AitSahalia2014}. Here, we focus in particular on stochastic dynamics in networks \cite{Barrat2008,Newman2010}. Some challenging examples are the thermal dynamics in spin glasses, avalanche dynamics, the dynamics of infectious diseases, the evolution of opinions in social networks, the stability of functionality under perturbations in technological networks, or synchronization phenomena. The probabilistic features can be intrinsic or due to our ignorance of certain details that are not essential for the observed (macroscopic) phenomena. A simple method for the simulation of a stochastic system with states $\sigma_i$ on $N$ vertices of a network is to propagate the probability distribution $P(\sigma_1^t,\dotsc,\sigma_N^t)$ in time. Unfortunately, the corresponding computation costs are exponential in the system size $N$ and it is also difficult to asses temporal correlations in this way.

For the investigation of network systems, a strong simplification can be achieved when cycles in the interaction graph are rare or sufficiently long. This is the case for locally tree-like graphs [see Fig.~\ref{fig:graphicalRep}(a)] such as random regular graphs, Erd\H{o}s-R\'{e}nyi graphs, and Gilbert graphs. For such random graphs with $N$ vertices, almost all cycles have length $\gtrsim\log N$ such that their effect is negligible in the thermodynamic limit \cite{Mezard2009}. For static problems, this feature is used in the \emph{cavity method} \cite{Mezard1986-1,Mezard2001-20}, where conditional nearest-neighbor probabilities are computed iteratively within the Bethe-Peierls approximation. This efficient method has been applied very successfully to study, for example, equilibrium properties of spin glasses \cite{Mezard2001-20}, computationally hard satisfiability problems \cite{Mezard2002-297,Mezard2002-66}, and random matrix ensembles \cite{Rogers2008-78}. 

Subsequently, the cavity method has been generalized to dynamical problems, resulting in the \emph{dynamic cavity method} or dynamic belief propagation \cite{Neri2009-08,Karrer2010-82}. The central objects in this approach are not the time-dependent probabilities for global system states, but conditional probabilities $\mu_{i\to j}(\sigma^0_i,\dotsc,\sigma^t_i|\sigma^0_j,\dotsc,\sigma^{t-1}_j)$ for state trajectories on neighboring sites $i$ and $j$ (edges $i\to j$). These so-called edge messages $\mu_{i\to j}$ are generated in an iteration over time. Consequently, the computational complexity is now linear instead of exponential in the system size $N$. Unfortunately, the number of possible trajectories and, hence, the computational complexity increase exponentially in time. Applications have thus been restricted to either very short times \cite{Neri2009-08,Kanoria2011-21}, oriented graphs \cite{Neri2009-08}, a no-backtracking approximation \cite{Shrestha2015-92,Castellano2018-98}, or unidirectional dynamics with local absorbing states \cite{Karrer2010-82,Lokhov2015-91,Lokhov2014-90,Altarelli2014-112,Altarelli2014-10,Shrestha2014-89,Lokhov2017-114}. In the latter case, one can exploit that vertex trajectories can be parametrized by a few switching times.

For general stochastic network dynamics, it is a challenging endeavor to find good approximative solutions to the dynamic cavity equations with polynomial computation costs. A drastic simplification is to neglect temporal correlations completely as in the one-time approximation (a.k.a.\ time factorization) \cite{Neri2009-08,Aurell2011-04,Aurell2012-85,Zhang2012-148} or to retain only short-time correlations as in the one-step Markov ansatz \cite{DelFerraro2015-92}. While this can be expected to work well for stationary states at high temperatures, such approximations are usually severe for short to intermediate times or low temperatures. Other approximative approaches are the cluster variational method \cite{Pelizzola2013-86,Vazquez2017-3,Pelizzola2017-7} (applicable for short-range spatio-temporal correlations) or perturbative schemes \cite{Roudi2011,Aurell2012-85,BachschmidRomano2016-49}, generating functional analysis \cite{Hatchett2004-37,Mimura2009-42,Mozeika2011-106,Mozeika2012-92,Coolen2012-92}, and the generalized mean field approximation \cite{Mahmoudi2014-7}.

In Ref.~\cite{Barthel2018-97}, we have introduced a novel efficient algorithm for precise solutions of the parallel dynamic cavity equations for locally tree-like graphs and general bidirectional dynamics. The approach is based on \emph{matrix-product approximations} for the edge messages $\mu_{i\to j}$. The computational complexity is decreased from exponential to polynomial in the duration of the dynamical process. As we demonstrated for non-equilibrium Glauber dynamics in the kinetic Ising model, one can obtain quasi-exact results with relatively small matrix dimensions. Computation costs and accuracy can be tuned through controlled truncations of the matrix dimensions. Matrix-product approximations are widely used in condensed matter theory to encode spatial correlations of quantum states \cite{Accardi1981,Fannes1992-144,Rommer1997,Schollwoeck2011-326}. Here, we employ matrix products to encode temporal correlations in the edge messages.

This paper gives a pedagogical introduction to the dynamic cavity method in Sec.~\ref{sec:stochDyn}. Sections~\ref{sec:mpem}-\ref{sec:mpemExpect} give a detailed account of the novel matrix product edge message (MPEM) algorithm \cite{Barthel2018-97} for the solution of the dynamic cavity equation. Section~\ref{sec:dmTrunc} provides an improved MPEM truncation scheme that substantially reduces computation costs. For Glauber-Ising dynamics, we demonstrate a linear growth of computation costs in Sec.~\ref{sec:example}. In Sec.~\ref{sec:continuum}, the MPEM method is generalized to systems where the transition probability for the state of a vertex does not only depend on the state of its neighbors in the interaction graph, but also on the current state of the vertex itself.

\section{Stochastic dynamics on locally tree-like graphs} \label{sec:stochDyn}
\vspace{-0.2em}
\subsection{Global equation of motion}
Let us consider a graph $\G$ and denote the state at vertex $i\in\G$ by $\sigma_i$. The state of the full system at time $t$ will be denoted by
\begin{equation*}
	\vs^t:=(\sigma^t_1,\sigma^t_2,\dots).
\end{equation*}
Given the state probabilities $P(\vs^t)$ for time $t$, the probabilities $P(\vs^{t+1})$ for the subsequent time step $t+1$ are obtained by applying the transition matrix $W(\vs^{t+1}|\vs^t)$, where the notation of the arguments indicates that $W$ is a conditional probability.
\begin{equation}\label{eq:eom_global}
	P(\vs^{t+1}) = \sum_{\vs^t} W(\vs^{t+1}|\vs^t) P(\vs^t)
	\quad\text{with}\quad \sum_{\vs} W(\vs|\vs')=1.
\end{equation}
In correspondence with the structure of the graph $\G$, we assume that, in every time step, the probability for $\sigma_i^{t+1}$ only depends on the states $\sigma_k^{t}$ of nearest-neighbors $k$ at the previous time step such that the global transition matrix $W$ is a product of local transition matrices $w_i$,
\begin{equation}\label{eq:transitionMatrix}
	W(\vs^{t+1}|\vs^t) = \prod_{i\in \G} w_i(\sigma_i^{t+1}|\vs_{\partial i}^{t})
	\quad\text{with}\quad \sum_{\sigma_i} w_i(\sigma_i|\vs'_{\partial i})=1,
\end{equation}
where the vicinity $\partial i:=\{k\in\G\,|\,\dist(i,k)\leq 1\}$ of vertex $i$, contains $i$ and its nearest neighbors. Equations \eqref{eq:eom_global} and \eqref{eq:transitionMatrix} specify the parallel stochastic dynamics of the system.

\subsection{Probability for global trajectories}
With the definitions above, the probability for a trajectory $\vs^0\to\vs^1\to\dots\to\vs^t$ of states up to time $t$ takes the form
\begin{equation}
	P(\vs^0\vs^1\dots\vs^{t-1}\vs^t)=\prod_{s=1}^t W(\vs^s|\vs^{s-1})P(\vs^0).
\end{equation}
For simplicity, let us assume initially uncorrelated states, i.e.,
\begin{equation}
	P(\vs^0)=\prod_{i\in\G}p_i(\sigma_i^0)
	\quad\text{with}\quad \sum_{\sigma_i} p_i(\sigma_i)=1.
\end{equation}

\subsection{Motivation and definition of edge messages} \label{sec:em_def}
\begin{figure}[b]
\label{fig:graphicalRep}
\includegraphics[width=\textwidth]{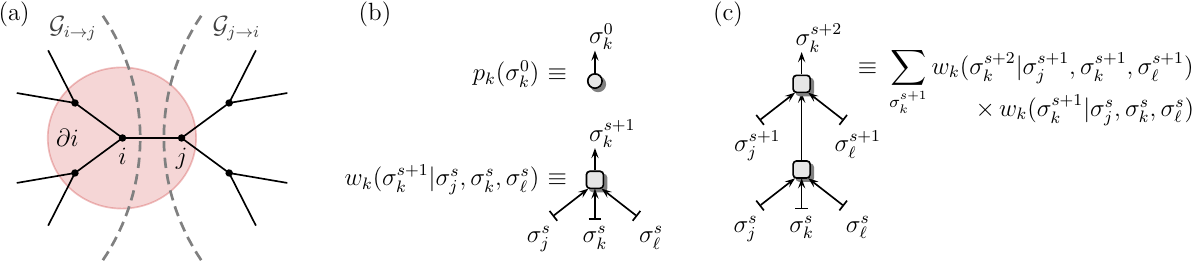}
\caption{(a) Example for subgraphs $\G_{i\to j}$ and $\G_{j\to i}$ as defined in Sec.~\ref{sec:em_def}. (b) Graphical representations for single-vertex probabilities $p_k(\sigma_k)$ and local transition matrices $w_k(\sigma_k^{s+1}|\vs_{\partial k}^s)$ for the case of vertex degree 3 ($\partial k=\{j,k,\ell\}$). (c) Graphical representation for the (partial) contraction of two transition matrices.}
\end{figure}
An exact solution of the global equation of motion \eqref{eq:eom_global} is usually not possible. In principle, one can employ a simple iteration over time. However, the corresponding memory costs (to keep track of $P(\vs^t)$) and computation costs scale \emph{exponentially} in the system size $|\G|$ such that this approach is limited to very small systems.

A very useful technique is the Markov chain Monte Carlo method. In this approach, one repeatedly generates states $\vs^0$ according to the $t=0$ probability $P(\vs^0)$, and propagates them in time by choosing a state $\vs^{t+1}$ with probability $W(\vs^{t+1}|\vs^{t})$ for $t=0,1,2,\dotsc$. While the technique can be used to study many interesting problems, it has several disadvantages. First of all, one cannot address the thermodynamic limit directly and needs to perform a costly finite-size scaling analysis. Secondly, the number of required samples can be very large if temporal correlations are non-trivial. Also, the error scaling is not very favorable. When increasing the number of samples $N_s$, errors decrease slowly as $1/\sqrt{N_s}$. Especially if the absolute value of the desired observable is small, as is usually the case when studying temporal correlations, it is often impossible to achieve a sufficient precision.

To resolve these issues in our approach for locally tree-like graphs, we shift the focus from global-state probabilities $P(\vs^t)$ to so-called edge messages. These are conditional probabilities for the \emph{trajectory} $\bs^{t+1}_i$ of the variable at vertex $i$,
\begin{equation}
	\bs^{t+1}_i:=(\sigma^0_i,\sigma^1_i,\dotsc,\sigma^{t}_i,\sigma^{t+1}_i),
\end{equation}
and the trajectory $\bs^{t}_j$ on a neighboring site $j$. If the graph is a tree, the exact evolution of the system can be formulated in terms of the edge messages and leads to the dynamic cavity equations \cite{Neri2009-08,Karrer2010-82} discussed in the following. If the graph contains some longer loops, i.e., is only locally tree-like, the dynamic cavity equations give an approximation to the exact dynamics with the accuracy depending on the number and lengths of loops.
\begin{figure}[t]
\label{fig:edgeMsg_1D}
\includegraphics[width=0.58\textwidth]{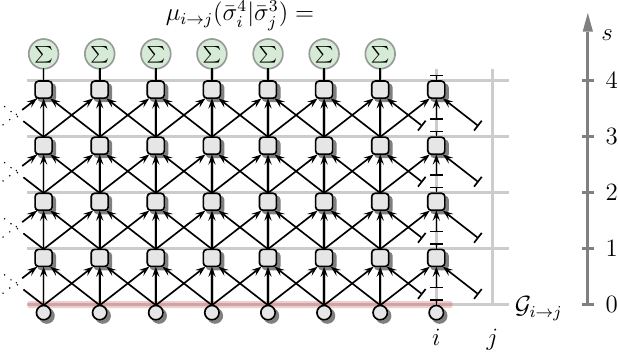}
\caption{Edge message $\mu_{i\to j}(\bs_i^4|\bs_j^3)$, as defined in Eq.~\eqref{eq:edgeMessage}, for a one-dimensional graph.}
\end{figure}

When we remove an edge $(i,j)$ from a tree graph $\G$, it is decomposed into two parts $\G_{i\to j}$ and $\G_{j\to i}$ such that $\G_{i\to j}$ contains $i$ and the subgraph that is still connected to $i$ (and all subgraphs of $\G$ that are disconnected from $i$ and $j$) while $\G_{j\to i}$ contains $j$ and the subgraph that is still connected to $j$. See Fig.~\ref{fig:graphicalRep}(a). Now, an \emph{edge message} $\mu_{i\to j}(\bs_i^{t+1}|\bs_j^{t})$ for edge $(i,j)$ is the probability of the trajectory $\bs_i^{t+1}$ on vertex $i$ for the dynamics being restricted to the subgraph $\G_{i\to j}$ under the condition that we impose the trajectory $\bs_j^{t}$ on vertex $j$. Specifically, it is the product of all transition matrices $w_k(\sigma_k^{s+1}|\vs_{\partial k}^s)$ and $p_k(\sigma_k^0)$ for all $k\in\G_{i\to j}$ and $0\leq s\leq t$, summed over all $\bs_k^{t+1}$ with $k\in\G_{i\to j}\setminus\{i\}$, i.e.,
\begin{equation}\label{eq:edgeMessage}
	\mu_{i\to j}(\bs_i^{t+1}|\bs_j^{t}):=\sum_{\{\bs_k^{t+1}\}_{k\in\G_{i\to j}\setminus\{i\}}}
	  \Big[\prod_{s=0}^{t}\prod_{q\in\G_{i\to j}}w_q(\sigma_q^{s+1}|\vs_{\partial q}^{s})\Big]
	  \prod_{q\in\G_{i\to j}}p_q(\sigma_q^0).
\end{equation}
A pictorial representation for a one-dimensional graph is given in Fig.~\ref{fig:edgeMsg_1D}.

\subsection{Simplification of edge messages}
\begin{figure}[p]
\label{fig:edgeMsg_simplified_1D}
\includegraphics[width=0.87\textwidth]{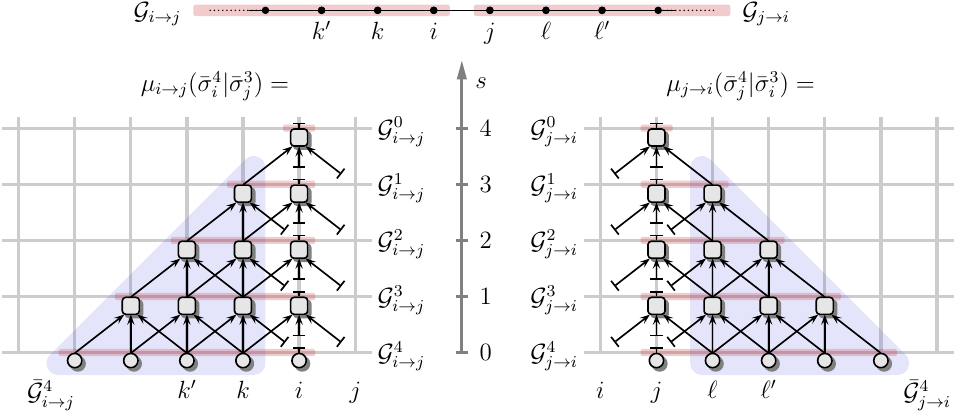}
\caption{Simplified forms of edge messages $\mu_{i\to j}(\bs_i^4|\bs_j^3)$ and $\mu_{j\to i}(\bs_j^4|\bs_i^3)$ for a one-dimensional graph, where, as in Eq.~\eqref{eq:edgeMessageC}, the normalization constraints of transition matrices $w_k$ and $t=0$ probabilities $p_k(\sigma_k^0)$ have been used to remove those outside the causal cones $\bG^4_{i\to j}$ and $\bG^4_{j\to i}$. The latter are defined in Eq.~\eqref{eq:GC} and are built from subgraphs $\G^s_{i\to j}$ and $\G^s_{j\to i}$, respectively.}
\end{figure}

\begin{figure}[p]
\label{fig:edgeMsg_simplified_Y}
\includegraphics[width=\textwidth]{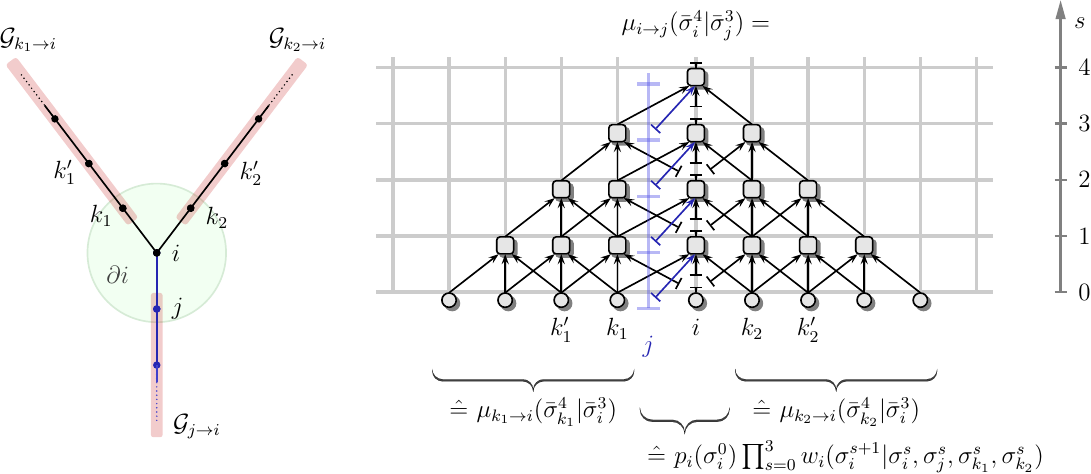}
\caption{Simplified form \eqref{eq:edgeMessageC} of the edge message $\mu_{i\to j}(\bs_i^4|\bs_j^3)$ for a Y-junction graph with central vertex $i$, where normalization constraints have been used to remove transition matrices $w_k$ and $t=0$ probabilities $p_k(\sigma_k^0)$ outside the causal cone $\bG^4_{i\to j}$. As shown here and specified in Eq.~\eqref{eq:dynamicCavityEq}, the edge message $\mu_{i\to j}$ for time $t+1$ can be obtained by contracting $\mu_{k\to i}$ for time $t$ and all vertices $k\neq i,j$ in the neighborhood $\partial i$ with transition matrices $w_i(\sigma_i^{s+1}|\vs_{\partial i}^s)$ and $p_i(\sigma_i^0)$.}
\end{figure}

Due to the normalization constraint $\sum_{\sigma_i} w_i(\sigma_i|\vs'_{\partial i})=1$ of the transition matrices, the expression \eqref{eq:edgeMessage} for the edge message can be simplified substantially by removing all $w_k$ and $p_k$ outside the \emph{causal cone} of edge $i\to j$ such that
\begin{equation}\label{eq:edgeMessageC}
	\mu_{i\to j}(\bs_i^{t+1}|\bs_j^{t})=\sum_{\{\sigma_k^s\}_{(k,s)\in\bG^{t+1}_{i\to j}}}
	  \Big[\prod_{s=0}^{t}\prod_{q\in\G^{t-s}_{i\to j}}w_q(\sigma_q^{s+1}|\vs_{\partial q}^{s})\Big]
	  \prod_{q\in\G^{t+1}_{i\to j}}p_q(\sigma_q^0).
\end{equation}
In this equation, we use the (recursively defined) subgraphs
\begin{equation}\label{eq:Gsub}
	\G^0_{i\to j}:=\{i\},\quad 
	\G^{s+1}_{i\to j}:=\G^{s}_{i\to j}\cup\Big[\bigcup_{k\in \G^{s}_{i\to j}}(\partial k\setminus\{j\})\Big]
	 = \big\{k\in \G_{i\to j}\,|\,\dist(i,k)\leq s+1\big\}
\end{equation}
and the \emph{causal cone} $\bG^{t+1}_{i\to j}$ of edge $i\to j$ which is a subset of $\mathbb{N}_0\times \G_{i\to j}$,
\begin{equation}\label{eq:GC}
	\bG^{t}_{i\to j}:=\bigcup_{s=0}^t\Big\{(s,k)\,|\, k\in \G^{t-s}_{i\to j}\setminus\{i\}\Big\}.
\end{equation}
See Figs.~\ref{fig:edgeMsg_simplified_1D} and \ref{fig:edgeMsg_simplified_Y} for examples.

\subsection{Computing edge trajectory probabilities from edge messages}\label{sec:em_observe}
\begin{figure}[t]
\label{fig:edgeHistoryProb_1D}
\includegraphics[width=0.7\textwidth]{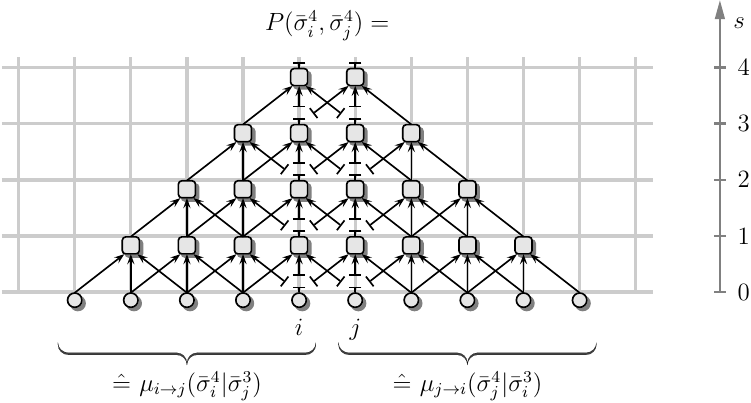}
\caption{The joint probability \eqref{eq:edgeProb} for the trajectories $\bs_i^4$ and $\bs_j^4$ on an edge $(i,j)$ is given by the product of the edge messages $\mu_{i\to j}(\bs_i^4|\bs_j^3)$ and $\mu_{j\to i}(\bs_j^4|\bs_i^3)$. It is shown here for a one-dimensional graph.}
\end{figure}

Before showing that edge messages can be generated in an iteration (Sec.~\ref{sec:em_iteration}), note that they allow us to evaluate easily many observables of interest. In particular, the product of edge messages $\mu_{i\to j}(\bs_i^t|\bs_j^{t-1})$ and $\mu_{j\to i}(\bs_j^t|\bs_i^{t-1})$ yields the joint probability for the occurrence of trajectories $\bs_i^t$ and $\bs_j^t$ on vertices $i$ and $j$,
\begin{equation}\label{eq:edgeProb}
	P(\bs_i^t,\bs_j^t) \equiv \sum_{\{\bs_k^t\}_{k\in \G\setminus\{i,j\}}} P(\vs^0\vs^1\dots\vs^{t-1}\vs^t)
	  = \mu_{i\to j}(\bs_i^t|\bs_j^{t-1}) \mu_{j\to i}(\bs_j^t|\bs_i^{t-1}).
\end{equation}
See Fig.~\ref{fig:edgeHistoryProb_1D}.

By marginalizing Eq.~\eqref{eq:edgeProb} over certain subsets of the variables $\{\sigma_i^s\}_{0\leq s\leq t}$ and $\{\sigma_j^s\}_{0\leq s\leq t}$, one obtains the probabilities required for the evaluation of time-local observables or temporal correlation functions.

\subsection{Dynamic cavity equation} \label{sec:em_iteration}
Inspection of the simplified expression \eqref{eq:edgeMessageC} for edge message $\mu_{i\to j}(\bs_i^{t+1}|\bs_j^{t})$ shows that it can be constructed from the time-$t$ edge messages $\mu_{k\to i}(\bs_k^{t}|\bs_i^{t-1})$ according to
\begin{equation}\label{eq:dynamicCavityEq}
	\mu_{i\to j}(\bs_i^{t+1}|\bs_j^{t}) = \sum_{\{\bs_k^t\}_{k\in \partial i\setminus\{i,j\}}}
	  p_i(\sigma_i^0) \Big[\prod_{s=0}^t w_i(\sigma_i^{s+1}|\vs_{\partial i}^s)\Big]
	  \Big[\prod_{k\in\partial i\setminus\{i,j\}} \mu_{k\to i}(\bs_k^{t}|\bs_i^{t-1})\Big]
\end{equation}
with $\mu_{i\to j}(\sigma_i^0)\equiv p_i(\sigma_i^0)$.
This recursion relation for the edge messages, the \emph{dynamic cavity equation} \cite{Neri2009-08,Karrer2010-82}, is very useful and central for the matrix product edge message time-evolution algorithm presented in Sec.~\ref{sec:mpemAlgo}. An illustration of Eq.~\eqref{eq:dynamicCavityEq} for the example of a Y-junction graph is given in Fig.~\ref{fig:edgeMsg_simplified_Y}.

\section{Matrix product edge messages (MPEM)} \label{sec:mpem}
This section and Sec.~\ref{sec:mpemAlgo} give a detailed account of an evolution algorithm that employs precise approximations to edge messages \eqref{eq:edgeMessageC} in matrix product form \cite{Barthel2018-97}. Extensions will be described in Secs.~\ref{sec:dmTrunc} and \ref{sec:continuum}.

\subsection{General idea and motivation}\label{sec:mpemMotivation}
In the following let $d$  denote the number of different vertex states ($\sigma_i=1,\dotsc,d$).
In Sec.~\ref{sec:stochDyn}, we have seen that the stochastic dynamics on a tree graph $\G$ can not only be simulated by sampling trajectories $(\vs^0\vs^1\dots\vs^{t-1}\vs^t)$ of the full system according to their weights \eqref{eq:eom_global} using a Monte Carlo algorithm, but also by an iterative construction of edge messages $\mu_{i\to j}(\bs_i^{t}|\bs_j^{t-1})$ which give access to many observables of interest such as the evolution of time-local observables or temporal correlators. Importantly, the computation costs and memory requirements for evolving and storing an edge message ($\mc{O}(d^{2t-1})$) are independent of the system size. However, without approximations, the costs increase exponentially with time $t$. So, an exact treatment is limited to short times.

To resolve this problem, we can employ an idea from the \emph{density matrix renormalization group} (DMRG) \cite{White1992-11,White1993-10,Schollwoeck2005} which is a numerical algorithm for the simulation of strongly-correlated quantum many-body systems (predominantly for one-dimensional lattices) which has been used very successfully for a wide range of applications in condensed matter physics, the physics of ultracold atomic gases, and quantum chemistry. In principle, the encoding of a many-body state for a lattice of $L$ sites,
\begin{equation}
	|\psi\ket=\sum_{n_1,\dotsc,n_L}\psi_{n_1,\dotsc,n_L}|\vec{n}\ket,
	\quad\text{where}\quad
	|\vec{n}\ket\equiv|n_1\ket\otimes|n_2\ket\otimes\dotsc\otimes|n_L\ket
\end{equation}
label orthonormal basis states ($\bra\vec{n}|\vec{n}'\ket=\delta_{\vec{n}\vec{n}'}$), requires storage of the $d^L$ expansion coefficients $\psi_{n_1,\dotsc,n_L}$. However, in ground states of typical quantum many-body systems, spatial correlations like $\bra\hat{S}^+_x\hat{S}^-_{x'}\ket$ decay quickly in the distance $|x-x'|$, exponentially or algebraically. Due to this fact and corresponding entanglement properties, one can approximate $\psi_{n_1,\dotsc,n_L}$ by much fewer effective degrees of freedom, given by the elements of $M_x\times M_{x+1}$ matrices $A_x^{n_x}$ in the approximation \cite{Hastings2007-76,Brandao2013-9,Verstraete2005-5,Barthel2017_08unused}
\begin{equation}\label{eq:MPSappr}
	|\psi\ket \approx \sum_{\vec{n}} A_1^{n_1}A_2^{n_1}\dotsb A_{L-1}^{n_{L-1}}A_L^{n_L}\,|\vec{n}\ket.
\end{equation}
The $M_x$ are called \emph{bond dimensions}. In order for the matrix product $A_1^{n_1}\dotsb A_L^{n_L}$ to yield a scalar, we require dimensions $M_1=M_{L+1}=1$ at the system boundaries. The elements of matrices $A_x^{n_x}$ are effective degrees of freedom that encode correlations around site $x$. The larger the bond dimensions $M_x$ are chosen, the more effective degrees of freedom are taken into account, the higher the computation costs, and the higher the precision of the approximation. In any case, the memory costs ($\mc{O}(dLM^2)$) and computation costs ($\mc{O}(d^kLM^3)$) are now linear in the system size $L$. The right-hand side of Eq.~\eqref{eq:MPSappr} is called a \emph{matrix product state} (MPS) \cite{Accardi1981,Fannes1992-144,Rommer1997,Schollwoeck2011-326}. More recently, MPS are also discussed in the more mathematical literature, sometimes under the name \emph{tensor train decomposition} \cite{Oseledets2011-33}. DMRG is a class of algorithms operating on MPS -- most importantly, to compute ground state approximations, thermal equilibrium states, or to study non-equilibrium phenomena. Observables can often be computed to machine precision.

The idea is now to similarly exploit that (connected) \emph{temporal} correlations in edge messages often decay quickly in time $t$ and/or in the time difference $|t-t'|$ and to approximate edge messages $\mu_{i\to j}(\bs_i^{t}|\bs_j^{t-1})$ in the form of a matrix product.

\subsection{Canonical form}
\begin{figure}[b]
\label{fig:MPEM_canonical}
\includegraphics[width=0.78\textwidth]{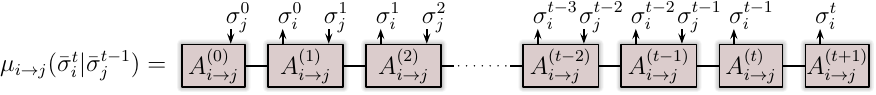}
\caption{Canonical form \eqref{eq:MPEM} of an MPEM $\mu_{i\to j}(\bs_i^{t}|\bs_j^{t-1})$. For every time step $s$ with $1\leq s\leq t-1$, there are $d^2$ matrices $A^{(s)}_{i\to j}(\sigma_i^{s-1}|\sigma_j^s)$ of dimension $M_s\times M_{s+1}$. With $M:=\max_s M_s$, the number of degrees of freedom is of order $td^2M^2$, i.e., linear in time instead of exponential in time.}
\end{figure}

Let us define the canonical form of a \emph{matrix product edge message} (MPEM) as
\begin{equation}\label{eq:MPEM}
	\mu_{i\to j}(\bs_i^{t}|\bs_j^{t-1}) = A^{(0)}_{i\to j}(\sigma_j^0) \Big[\prod_{s=1}^{t-1} A^{(s)}_{i\to j}(\sigma_i^{s-1}|\sigma_j^s)\Big] A^{(t)}_{i\to j}(\sigma_i^{t-1})A^{(t+1)}_{i\to j}(\sigma_i^{t})
\end{equation}
The graphical representation is shown in Fig.~\ref{fig:MPEM_canonical}. The particular choice of assigning state labels $\{\sigma^s_i\}$ and $\{\sigma^s_j\}$ to the $M_s\times M_{s+1}$ matrices occurring in the matrix product \eqref{eq:MPEM} is advantageous for the implementation of the dynamic cavity equation \eqref{eq:dynamicCavityEq} for MPEMs as will become clear in Sec.~\ref{sec:mpemAlgo}. In order for the matrix product \eqref{eq:MPEM} to yield a scalar, we set $M_0=M_{t+2}=1$.

\subsection{Controlled truncations}\label{sec:MPS_trunc}
When advancing an MPEM $\mu_{i\to j}$ one step in time according to the dynamic cavity equation \eqref{eq:dynamicCavityEq}, we need to contract \Footnote{With the \emph{contraction} of two tensors, we refer to their product, summed over certain sets of common indices.} several MPEMs, $\mu_{k\to i}(\bs_i^{t}|\bs_j^{t-1})$ with $k$ from the neighborhood $\partial i$ of vertex $i$, and local transition matrices $w_i$ and, subsequently, express the result again in MPEM form. In general, this generates an MPEM $\mu_{i\to j}(\bs_i^{t+1}|\bs_j^{t})$ with increased bond dimension $M'\geq M$. In order to control the growth of the bond dimensions, and hence the computation costs, we would like to reduce the bond dimension of the new MPEM. This can indeed be done in a controlled fashion such that the resulting truncated MPEM is close to the original. Here, we discuss the most simple truncation scheme \cite{Barthel2018-97} and an optimized scheme is presented in Sec.~\ref{sec:dmTrunc}. The iterative time evolution of MPEMs is described in Sec.~\ref{sec:mpemAlgo}.

Because the notation is a little simpler, let us demonstrate truncations using the example of an MPS for a quantum system of $L$ sites,
\begin{equation}\label{eq:MPS}
	|\psi\ket= \sum_{\vec{n}} A_1^{n_1}A_2^{n_1}\dotsb A_{L-1}^{n_{L-1}}A_L^{n_L}\,|\vec{n}\ket,
\end{equation}
Let us split the system into two parts $\mc{L}$, containing sites $[1,\ell]$, and $\mc{R}$, containing sites $[\ell+1,L]$. Let $\{|a\ket\}$ and $\{|b\ket\}$ be orthonormal bases for left and right parts, respectively, such that
\begin{equation}\label{eq:psiBeforeTrunc}
	|\psi\ket=:\sum_{a,b}\psi_{a,b}|a\ket\otimes|b\ket.
\end{equation}
The objective is to find a good approximation $|\psi_\trunc\ket$ of $|\psi\ket$ in a reduced vector space, where the 2-norm distance is used to quantify the accuracy. This can be done using a singular value decomposition (SVD) of $\psi_{a,b}$. In matrix form, it reads $\psi=U\Lambda V^\dag$, where $U$ and $V$ are unitary matrices and $\Lambda$ is the diagonal matrix containing descendingly ordered  \emph{singular values} $\lambda_1\geq\lambda_2\geq\dots\geq \lambda_{M'}$ such that
\begin{subequations}
\begin{equation}
	|\psi\ket=\sum_{a,b,k}U_{a,k} \lambda_k V^*_{b,k}|a\ket\otimes|b\ket
	 =\sum_{k=1}^{M'} \lambda_k|k_\mc{L}\ket\otimes|k_\mc{R}\ket,
\end{equation}
where the states $\{|k_\mc{L}\ket:=\sum_a U_{a,k}|a\ket\}$ and $\{|k_\mc{R}\ket:=\sum_b V^*_{b,k}|b\ket\}$ are also orthonormal bases for $\mc{L}$ and $\mc{R}$, respectively. The best approximation $|\psi\ket$ with  
$M<M'$ nonzero singular values (also known as Schmidt coefficients in this case) is given by
\begin{equation}\label{eq:psiTrunc}
	|\psi_\trunc\ket:=\sum_{k=1}^M \lambda_k|k_\mc{L}\ket\otimes|k_\mc{R}\ket
	\quad \text{with error}\quad
	\norm{\psi-\psi_\trunc}^2=\sum_{k=M+1}^{M'}\lambda_k^2
\end{equation}
\end{subequations}
The fact that this procedure yields the best rank-$M$ approximation of $|\psi\ket$ for the given bipartition corresponds to the Eckart-Young theorem \cite{Golub1996,Stewart1993-35}.

So, to truncate in this fashion bond dimensions of an MPS $|\psi\ket$ (and analogously for MPEM), we need to express it in suitable orthonormal bases for $\mc{L}$ and $\mc{R}$. This can be achieved by exploiting the freedom to replace two subsequent matrices $(A^{n_x}_x,A^{n_{x+1}}_{x+1})$ in the matrix product \eqref{eq:MPS} by $(A^{n_x}_x X^{-1},XA^{n_{x+1}}_{x+1})$. The inserted non-singular matrices $X$ and $X^{-1}$ clearly leave $|\psi\ket$ invariant as they cancel in the matrix product. Using this gauge freedom, we can bring the matrix product \eqref{eq:MPS} into the form
\begin{subequations}\label{eq:MPS-ON}
\begin{gather}
	|\psi\ket = \sum_{\vec{n}} Y_1^{n_1}\dotsb Y_\ell^{n_\ell}\cdot C\cdot Z_{\ell+1}^{n_{\ell+1}}\dotsb Z_L^{n_L}\,|\vec{n}\ket \quad\text{with} \\ \label{eq:ONconstr}
	\sum_n (Y_i^n)^\dag Y_i^n = \id\quad\text{and}\quad
	\sum_n Z_i^n (Z_i^n)^\dag = \id.
\end{gather}
\end{subequations}
The left and right \emph{orthonormality constraints} \eqref{eq:ONconstr} are imposed by sequences of singular value decompositions, sweeping from site 1 to site $\ell$ and from site $L$ to site $\ell+1$ with details discussed in Sec.~\ref{sec:B_from_C}.
The resulting matrix product \eqref{eq:MPS-ON} is in fact of the form \eqref{eq:psiBeforeTrunc}, with $\psi_{a,b}\leftrightarrow C_{a,b}$, and orthonormal basis states
\begin{subequations}
\begin{gather}
	|a\ket=\sum_{n_1,\dotsc,n_\ell} \big[Y_1^{n_1}\dotsb Y_\ell^{n_\ell}\big]_{1,a}\,|n_1\dots n_\ell\ket \quad \text{for} \ \mc{L}
	\ \text{and}\\
	|b\ket=\sum_{n_{\ell+1},\dotsc,n_L} \big[Z_{\ell+1}^{n_{\ell+1}}\dotsb Z_L^{n_L}\big]_{b,1}\,|n_{\ell+1}\dots n_L\ket \quad \text{for} \ \mc{R}.
\end{gather}
\end{subequations}
The orthonormality of these states follows from Eq.~\eqref{eq:ONconstr}: 
\begin{align}\nonumber
	\bra a'|a\ket&=\sum_{n_1,\dotsc,n_\ell}\big[(Y_\ell^{n_\ell})^\dag\dotsb (Y_3^{n_3})^\dag(Y_2^{n_2})^\dag(Y_1^{n_1})^\dag Y_1^{n_1}Y_2^{n_2}Y_3^{n_3}\dotsb Y_\ell^{n_\ell}\big]_{a',a}\\\label{eq:ONleft}
	 &=\sum_{n_2,\dotsc,n_\ell}\big[(Y_\ell^{n_\ell})^\dag\dotsb (Y_3^{n_3})^\dag(Y_2^{n_2})^\dag Y_2^{n_2}Y_3^{n_3}\dotsb Y_\ell^{n_\ell}\big]_{a',a}
	 =\dots = [\id]_{a',a}=\delta_{a,a'}
\end{align}
and similarly for the states $\{|b\ket\}$.
With a singular value decomposition of the matrix $C$, we then obtain an optimally truncated state $|\psi_\trunc\ket$ [Eq.~\eqref{eq:psiTrunc}] in MPS form.
\begin{figure}[b]
\label{fig:C_from_Aw_bulk}
\includegraphics[width=0.9\textwidth]{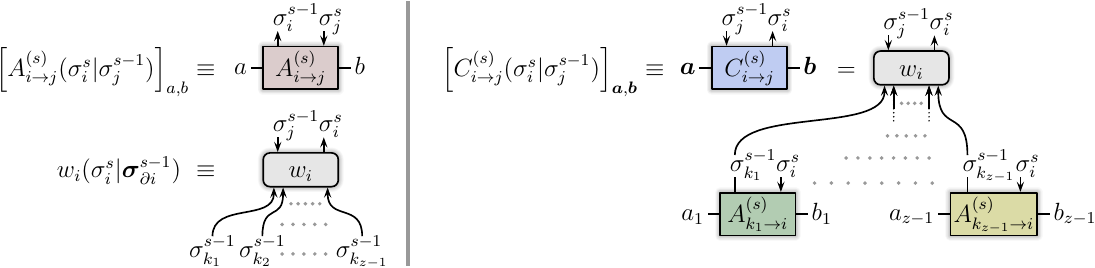}
\caption{Construction of tensors $C_{i\to j}^{(s)}$ for the evolved MPEM $\mu_{i\to j}(\bs_i^{t+1}|\bs_j^{t})$ of edge $i\to j$ at time $t+1$  using the local transition matrix $w_i(\sigma_i^{s}|\vs_{\partial i}^{s-1})$ and tensors $A_{k\to i}^{(s)}$ of time-$t$ MPEMs of neighboring edges $k\to i$. The resulting MPEM \eqref{eq:MPEM_C} is not yet in the canonical form \eqref{eq:MPEM_B}. The contraction shown here applies to the bulk $0<s<t$. The contractions at the boundaries $s=0$ and $s=t,t+1$ are shown in Fig.~\ref{fig:C_from_Aw_boundary}.}
\end{figure}

\section{Algorithm for the time evolution of matrix product edge messages}\label{sec:mpemAlgo}
Given MPEMs $\mu_{i\to j}(\bs_i^{t}|\bs_j^{t-1})$ [Eq.~\eqref{eq:MPEM}] for all edges at time $t$, we want to do one time step according to the dynamic cavity equation \eqref{eq:dynamicCavityEq} and obtain MPEM approximations 
\begin{equation}\label{eq:MPEM_B}
	\mu_{i\to j}(\bs_i^{t+1}|\bs_j^{t}) = B^{(0)}_{i\to j}(\sigma_j^0) \Big[\prod_{s=1}^{t} B^{(s)}_{i\to j}(\sigma_i^{s-1}|\sigma_j^s)\Big] B^{(t+1)}_{i\to j}(\sigma_i^{t})B^{(t+2)}_{i\to j}(\sigma_i^{t+1})
\end{equation}
for the edge messages at time $t+1$.

Let us assume for now that vertex $i$ is not member of its neighborhood $\partial i$, i.e., that the local transition matrix $w_i(\sigma_i^{s+1}|\vs_{\partial i}^{s})$ is independent of $\sigma_i^{s}$. The more general case can also be handled as well but requires a slightly more complicated algorithm. The corresponding extension is described in Sec.~\ref{sec:continuum}.

\subsection{One exact MPEM evolution step}\label{sec:C_from_Aw}
\begin{figure}[t]
\label{fig:C_from_Aw_boundary}
\includegraphics[width=0.82\textwidth]{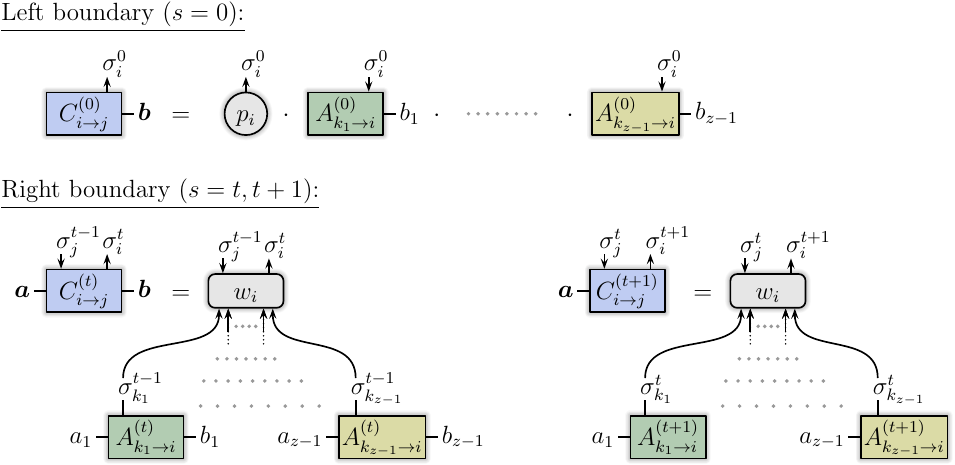}
\caption{Construction of the evolved MPEM $\mu_{i\to j}(\bs_i^{t+1}|\bs_j^{t})$ in the noncanonical form \eqref{eq:MPEM_C}, using transition matrices and MPEM tensors from the previous time step. The contractions shown here concern the boundaries $s=0$ and $s=t,t+1$. Those for the bulk $0<s<t$ are shown in Fig.~\ref{fig:C_from_Aw_bulk}.}
\end{figure}
First, we generate a non-canonical matrix product representation of the evolved edge message \eqref{eq:dynamicCavityEq}, in particular, choosing the non-canonical form
\begin{equation}\label{eq:MPEM_C}
	\mu_{i\to j}(\bs_i^{t+1}|\bs_j^{t}) = C^{(0)}_{i\to j}(\sigma_i^{0}) \Big[\prod_{s=1}^{t+1} C^{(s)}_{i\to j}(\sigma_i^{s}|\sigma_j^{s-1})\Big].
\end{equation}
To this purpose, the matrices $C^{(s)}_{i\to j}(\sigma_i^{s}|\sigma_j^{s-1})$ for the bulk, $0<s<t$, are obtained by contracting the transition matrix $w_i(\sigma_i^{s}|\vs_{\partial i}^{s-1})$ for vertex $i$ with the tensors $A^{(s)}_{k\to i}(\sigma_k^{s-1}|\sigma_i^s)$ from the time-$t$ MPEMs for edges $k\to i$ with vertices $k\in\partial i\setminus\{j\}=:\{k_1,\dotsc,k_{z-1}\}$, where $z$ is the degree of vertex $i$. In this contraction, we sum over the $z-1$ common indices $\sigma_k^{s-1}$ as illustrated in Fig.~\ref{fig:C_from_Aw_bulk}. The resulting matrices 
\begin{equation}\label{eq:C}
	C^{(s)}_{i\to j}(\sigma_i^{s}|\sigma_j^{s-1})
	= \sum_{\sigma_{k_1}^{s-1}\dotsc\sigma_{k_{z-1}}^{s-1}}
	    w_i(\sigma_i^{s}|\vs_{\partial i}^{s-1})
	  \Big[\bigotimes_{n=1}^{z-1} A^{(s)}_{k_n\to i}(\sigma_{k_n}^{s-1}|\sigma_i^s)\Big]
\end{equation}
have left and right multi-indices of dimensions $\bar{M}_s=(M_s)^{z-1}$ and $\bar{M}_{s+1}=(M_{s+1})^{z-1}$, corresponding to the direct products $\vec{a}:=(a_{k_1},\dotsc,a_{k_{z-1}})$ and $\vec{b}:=(b_{k_1},\dotsc,b_{k_{z-1}})$ of the left and right indices of the matrices $A^{(s)}_{k\to i}(\sigma_k^{s-1}|\sigma_i^s)$. For notational simplicity, we have assumed here that the bond dimensions $M_s$ are the same for all time-$t$ MPEMs of the edges $k\to i$ with $k\in\{k_1,\dotsc,k_{z-1}\}$.

Specifically, for the case $z=3$ with $\partial i=\{j,k_1,k_2\}$, the contractions take for example the form
\begin{multline*}
	\big[C^{(s)}_{i\to j}(\sigma_i^{s}|\sigma_j^{s-1})\big]_{(a_1,a_2),(b_1,b_2)} 
	  = \sum_{\sigma_{k_1}^{s-1},\sigma_{k_2}^{s-1}}
	    w_i(\sigma_i^{s}|\sigma_j^{s-1}\sigma_{k_1}^{s-1}\sigma_{k_2}^{s-1})\\[-0.3em]
	  \times\big[A^{(s)}_{k_1\to i}(\sigma_{k_1}^{s-1}|\sigma_i^s)\big]_{a_1,b_1} \cdot \big[A^{(s)}_{k_2\to i}(\sigma_{k_2}^{s-1}|\sigma_i^s)\big]_{a_2,b_2}.
\end{multline*}

At the extremal time slices $s=0$ and $s=t,t+1$, the contractions differ slightly. They are specified graphically in Fig.~\ref{fig:C_from_Aw_boundary}.

At this point, it has become clear why the specific choice for the assignment of state indices $\{\sigma_i^s\,|\,0\leq s\leq t\}$ and $\{\sigma_j^s\,|\,0\leq s\leq t-1\}$ to tensors $A_{i\to j}^{(s)}$ in the canonical MPEM form \eqref{eq:MPEM} is favorable. It allows for an entirely local construction of tensor $C_{i\to j}^{(s)}$ from tensors $A_{k\to i}^{(s)}$, i.e., only $A$-tensors of a single time-slice $s$ are involved.

\subsection{Truncating MPEMs and recovering the canonical form}\label{sec:B_from_C}
The exact progression by one time step yields the non-canonical MPEM \eqref{eq:MPEM_C} with bond dimensions being increased from $M_s$ to $\tilde{M}_s=(M_s)^{z-1}$. If we would proceed without any approximation, the computation cost would increase exponentially in time and, hence, restrict the simulation to very short times. Therefore, we are faced with two objectives: (a) reducing bond dimensions by a controlled truncation of the MPEM as described in Sec.~\ref{sec:MPS_trunc} and motivated physically in Sec.~\ref{sec:mpemMotivation}, and (b) rearranging the assignment of physical state indices $\{\sigma_i^s\}$ and $\{\sigma_j^s\}$ to attain the canonical form \eqref{eq:MPEM_B} of the evolved MPEM. This can be achieved in different ways: A relatively simple method \cite{Barthel2018-97} is described in the following and an optimized scheme is introduced in Sec.~\ref{sec:dmTrunc}.

If we want to truncate bond $s$, i.e., reduce the bond dimension $\tilde{M}_s$ to something smaller, we need to take care of orthonormality as discussed in Sec.~\ref{sec:MPS_trunc}. In particular, we need to express the edge message in orthonormal reduced bases for left block $[0,s]$ and right block $[s+1,t+1]$. This can be done by imposing left and right orthonormality constraints \eqref{eq:ONconstr} for the $C$-tensors. If we would not do so and truncate bond dimension $\tilde{M}_s$, say, through an SVD of tensor $C^{(s)}$ without having the other tensors in orthonormal form, the approximation error would be uncontrolled. Truncating the smallest singular values of $C^{(s)}$ would then not correspond to the best rank-$M_s$ approximation.
\begin{figure}[t]
\label{fig:MPEM_orthonormalize}
\includegraphics[width=\textwidth]{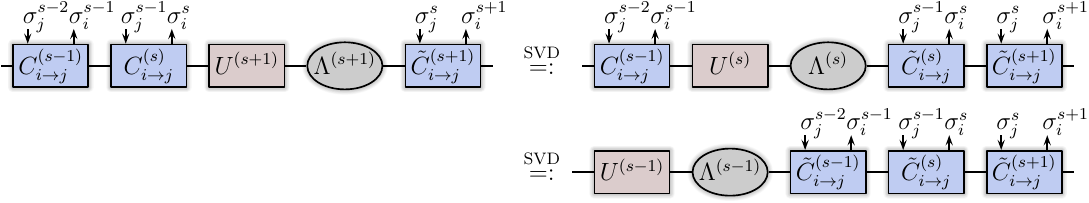}
\caption{In the truncation scheme described in Sec.~\ref{sec:B_from_C}, the evolved MPEM \eqref{eq:MPEM_C} needs to be orthonormalized before nonzero singular values can be truncated in a subsequent sweep. This orthonormalization is accomplished through a sequence of singular value decompositions, sweeping the matrix product from right to left. In the process, the $C$-tensors \eqref{eq:C} are replaced by $\tilde{C}$-tensors that obey the right orthonormality constraint; cf.\ Eq.~\eqref{eq:Cortho}.}
\end{figure}

In a first sweep from right ($s=t+1$) to left ($s=0$), we can sequentially impose the right orthonormality constraint $\sum_n Z_i^n (Z_i^n)^\dag = \id$ on all $C$-tensors. As the original $C$-tensors do not obey the orthonormality constraints, we can only truncate singular values $\lambda_k$ that are (up to machine precision) zero. At the right boundary, we start with the singular value decomposition $C^{(t+1)}_{i\to j}(\sigma_i^{t+1}|\sigma_j^{t})\stackrel{\SVD}{=:}U^{(t+1)}\Lambda^{(t+1)}\tilde{C}^{(t+1)}_{i\to j}(\sigma_i^{t+1}|\sigma_j^{t})$, continue for the bulk tensors $s=t,t-1,\dotsc,1$ with
\begin{equation}\label{eq:Cortho}
	C^{(s)}_{i\to j}(\sigma_i^{s}|\sigma_j^{s-1})U^{(s+1)}\Lambda^{(s+1)}\stackrel{\SVD}{=:}U^{(s)}\Lambda^{(s)}\tilde{C}^{(s)}_{i\to j}(\sigma_i^{s}|\sigma_j^{s-1}),
\end{equation}
as shown in Fig.~\ref{fig:MPEM_orthonormalize} and end at the left boundary with $\tilde{C}^{(0)}_{i\to j}(\sigma_i^{0}):=C^{(0)}_{i\to j}(\sigma_i^{0})U^{(1)}\Lambda^{(1)}$ such that all $\tilde{C}$-tensors except $\tilde{C}^{0}$ now obey the right orthonormality constraint. The computation cost for each such SVD is proportional to $\tilde{M}^3$.

In a subsequent sweep from left to right, again based on singular value decompositions, we can now do the actual truncations to decrease bond dimensions ($\tilde{C}\to\bar{C}$), e.g., by setting a threshold $\varepsilon$ and truncating all singular values $\lambda_k$ with $\lambda_k^2 <\varepsilon \cdot\sum_q\lambda_q^2$. All $\bar{C}$-tensors except $\bar{C}^{(t+1)}$ now obey the left orthonormality constraint and
\begin{equation}\label{eq:MPEM_Ctrunc}
	\mu_{i\to j}(\bs_i^{t+1}|\bs_j^{t})
	= \tilde{C}^{(0)}_{i\to j}(\sigma_i^{0}) \Big[\prod_{s=1}^{t+1} \tilde{C}^{(s)}_{i\to j}(\sigma_i^{s}|\sigma_j^{s-1})\Big]
	\stackrel{\trunc}{\approx}
	 \bar{C}^{(0)}_{i\to j}(\sigma_i^{0}) \Big[\prod_{s=1}^{t+1} \bar{C}^{(s)}_{i\to j}(\sigma_i^{s}|\sigma_j^{s-1})\Big].
\end{equation}
\begin{figure}[t]
\label{fig:MPEM_forms}
\includegraphics[width=0.9\textwidth]{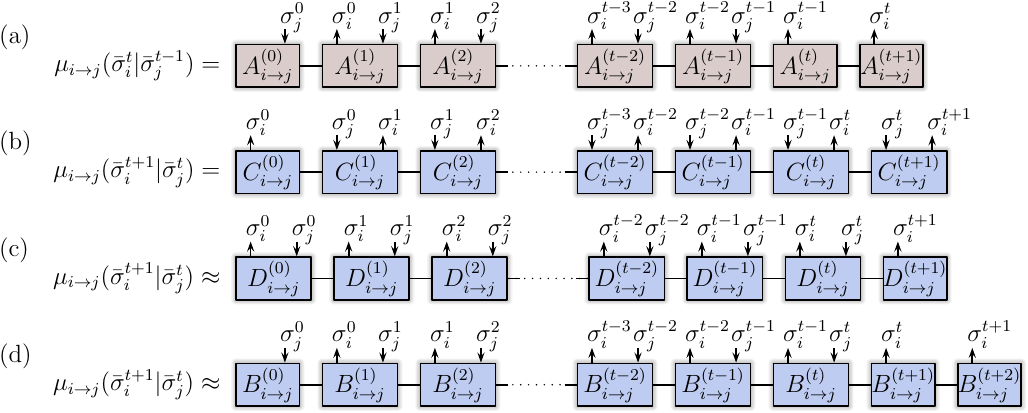}
\caption{Different forms of MPEMs during a time evolution step $t\to t+1$. (a) Canonical form \eqref{eq:MPEM} of MPEMs before the evolution step. (b) Non-canonical but exact form \eqref{eq:MPEM_C} of an evolved MPEM. (c) Non-canonical and approximative form \eqref{eq:MPEM_D} of the evolved MPEM after a sequence of truncations and reorderings of vertex variables. (d) Canonical approximative form \eqref{eq:MPEM_B} of the evolved MPEM after another sequence of truncations and vertex variable reorderings.}
\end{figure}

We now need to reorder the indices $\{\sigma_i^s\}$ and $\{\sigma_j^s\}$ of the vertex states. In a sweep from right to left, we go from the index assignment $(\sigma_i^{0})(\sigma_i^{1}|\sigma_j^{0})(\sigma_i^{2}|\sigma_j^{1})\dots(\sigma_i^{t+1}|\sigma_j^{t})$ in the truncated and orthonormalized version of the matrix product in Eq.~\eqref{eq:MPEM_Ctrunc} to the assignment $(\sigma_i^{0}\sigma_j^{0})(\sigma_i^{1}|\sigma_j^{1})\dots(\sigma_i^{t}|\sigma_j^{t})(\sigma_i^{t+1})$ in the matrix product
\begin{equation}\label{eq:MPEM_D}
	\mu_{i\to j}(\bs_i^{t+1}|\bs_j^{t}) \stackrel{\trunc}{\approx} \Big[\prod_{s=0}^{t} D^{(s)}_{i\to j}(\sigma_i^{s}|\sigma_j^{s})\Big] D^{(t+1)}_{i\to j}(\sigma_i^{t+1}).
\end{equation}
 At the right boundary, we start with a singular value decomposition and controlled truncation (SVT) $\bar{C}^{(t+1)}_{i\to j}(\sigma_i^{t+1}|\sigma_j^{t})\stackrel{\SVT}{\approx:}U^{(t+1)}(\sigma_j^{t})\Lambda^{(t+1)}D^{(t+1)}_{i\to j}(\sigma_i^{t+1})$ as described in footnote \Footnote{The resulting matrices $\mc{U}^\dag$ and $\mc{D}$, defined as $\mc{U}_{(a,\sigma),k}:=[U^{(t+1)}(\sigma)]_{a,k}$ and $\mc{D}_{k,(\sigma,b)}:=[D^{(t+1)}_{i\to j}(\sigma)]_{k,b}$ are isometric in the sense that $\mc{U}^\dag\mc{U}=\id =\mc{D}\mc{D}^\dag$. These matrices are obtained from the singular value decomposition $\bar{\mc{C}}\stackrel{\SVT}{\approx}\mc{U}\Lambda\mc{D}$ and subsequent truncation of small singular values $\lambda_k$, where $\bar{\mc{C}}_{(a,\sigma),(b,\sigma')}:=[\bar{C}^{(t+1)}_{i\to j}(\sigma|\sigma')]_{a,b}$. The isometric property of $\mc{D}$, corresponds to the right orthonormality constraint as defined in the second part of Eq.~\eqref{eq:ONconstr}.}, continue for the bulk tensors $s=t,t-1,\dotsc,1$ with
\begin{equation}
	\bar{C}^{(s)}_{i\to j}(\sigma_i^{s}|\sigma_j^{s-1})U^{(s+1)}(\sigma_j^{s})\Lambda^{(s+1)}\stackrel{\SVT}{\approx:}U^{(s)}(\sigma_j^{s-1})\Lambda^{(s)}D^{(s)}_{i\to j}(\sigma_i^{s}|\sigma_j^{s}),
\end{equation}
and end at the left boundary with $D^{(0)}_{i\to j}(\sigma_i^{0}|\sigma_j^{0}):=\bar{C}^{(0)}_{i\to j}(\sigma_i^{0})U^{(1)}(\sigma_j^{0})\Lambda^{(1)}$.

To finally obtain the evolved edge message in canonical form \eqref{eq:MPEM_B}, in a sweep from left to right, we go from the index assignment $(\sigma_i^{0}\sigma_j^{0})(\sigma_i^{1}|\sigma_j^{1})\dots(\sigma_i^{t}|\sigma_j^{t})(\sigma_i^{t+1})$ in Eq.~\eqref{eq:MPEM_D} to the canonical assignment $(\sigma_j^{0})(\sigma_i^{0}|\sigma_j^{1})(\sigma_i^{1}|\sigma_j^{2})\dots(\sigma_i^{t-1}|\sigma_j^{t})(\sigma_i^{t})(\sigma_i^{t+1})$.
At the left boundary, we start with $D^{(0)}_{i\to j}(\sigma_i^{0}|\sigma_j^{0})\stackrel{\SVT}{\approx:}B^{(0)}_{i\to j}(\sigma_j^{0})\Lambda^{(0)}V^{(0)}(\sigma_i^{0})$, continue for the bulk tensors $s=1,2,\dotsc,t$ with
\begin{equation}
	\Lambda^{(s-1)}V^{(s-1)}(\sigma_i^{s-1})D^{(s)}_{i\to j}(\sigma_i^{s}|\sigma_j^{s})\stackrel{\SVT}{\approx:}B^{(s)}_{i\to j}(\sigma_i^{s-1}|\sigma_j^{s})\Lambda^{(s)}V^{(s)}(\sigma_i^{s}),
\end{equation}
and end at the left boundary with $\Lambda^{(t)}V^{(t)}(\sigma_i^{t})D^{(t+1)}_{i\to j}(\sigma_i^{t+1})\stackrel{\SVT}{\approx:}B^{(t+1)}_{i\to j}(\sigma_i^{t})\Lambda^{(t+1)}V^{(t+1)}(\sigma_i^{t+1})$ and $B^{(t+2)}_{i\to j}(\sigma_i^{t+1}):=\Lambda^{(t+1)}V^{(t+1)}(\sigma_i^{t+1})$.
The matrix products for the different stages in the evolution and truncation of the edge message $\mu_{i\to j}$ are illustrated in Fig.~\ref{fig:MPEM_forms}.

\section{Evaluation of observables}\label{sec:mpemExpect}
\begin{figure}[t]
\label{fig:observe}
\includegraphics[width=0.87\textwidth]{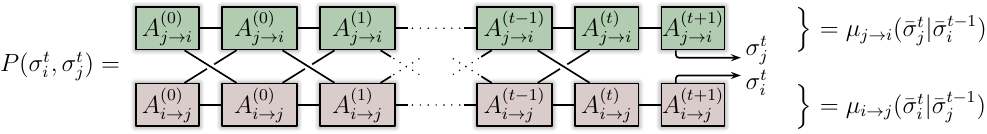}
\caption{Given MPEMs $\mu_{i\to j}(\bs_i^t|\bs_j^{t-1})$ and $\mu_{j\to i}(\bs_j^t|\bs_i^{t-1})$ for the edge $(i,j)$, observables such as the probability $P(\sigma_i^t,\sigma_j^t)$ of the edge being in state $(\sigma_i^t,\sigma_j^t)$ at time $t$ can be computed efficiently by (partial) contraction.}
\end{figure}

As described in Sec.~\ref{sec:em_observe}, the joint probability of trajectories $\bs_i^t$ and $\bs_j^{t-1}$ for the vertices of an edge $(i,j)$ is given by the product of the two corresponding edge messages, $P(\bs_i^t,\bs_j^t)= \mu_{i\to j}(\bs_i^t|\bs_j^{t-1}) \mu_{j\to i}(\bs_j^t|\bs_i^{t-1})$. Given matrix product representations (approximations) of these edge messages in canonical form \eqref{eq:MPEM}, time-local observables and temporal correlation functions can be evaluated efficiently. In order to evaluate, for example, the probability $P(\sigma_i^t,\sigma_j^t)$ of the edge being in state $(\sigma_i^t,\sigma_j^t)$ at time $t$, we simply contract all indices that occur in both MPEMs as depicted in Fig.~\ref{fig:observe}. The contractions can be started at the left boundary ($s=0$) in such a way that the total computation cost scales as $\mc{O}(tM^3)$.

To this purpose we start at the left boundary, i.e., at time slice $s=0$, with the multiplication $E^{(0)}(\sigma_i^0,\sigma_j^0):=A^{(0)}_{i\to j}(\sigma_j^0)\big[A^{(0)}_{j\to i}(\sigma_i^0)\big]^\intercal$, continue for the bulk tensors $s=0,1,\dotsc,t-2$ with
\begin{equation}\label{eq:mpem_observe}
	E^{(s+1)}(\sigma_i^{s+1},\sigma_j^{s+1}):=\sum_{\sigma_i^s} A^{(s+1)}_{i\to j}(\sigma_i^{s}|\sigma_j^{s+1}) \Big(\sum_{\sigma_j^s} E^{(s)}(\sigma_i^s,\sigma_j^s) \big[A^{(s+1)}_{j\to i}(\sigma_j^{s}|\sigma_i^{s+1})\big]^\intercal\Big),
\end{equation}
and finish with $E^{(t)}:=\sum_{\sigma_i^{t-1}} A^{(t)}_{i\to j}(\sigma_i^{t-1}) \big(\sum_{\sigma_j^{t-1}} E^{(t-1)}(\sigma_i^{t-1},\sigma_j^{t-1}) \big[A^{(t)}_{j\to i}(\sigma_j^{t-1})\big]^\intercal\big)$ and the final step $E^{(t+1)}(\sigma_i^{t},\sigma_j^{t}):= A^{(t+1)}_{i\to j}(\sigma_i^{t}) \big( E^{(t)} \big[A^{(t+1)}_{j\to i}(\sigma_j^{t})\big]^\intercal\big)$. Now, $E^{(t+1)}(\sigma,\sigma')$ are scalars ($1\times 1$ matrices because $M_{t+2}=1$) that yield the desired joint probability
\begin{equation}
	P(\sigma_i^t,\sigma_j^t) = E^{(t+1)}(\sigma_i^{t},\sigma_j^{t}) / \sum_{\sigma,\sigma'}E^{(t+1)}(\sigma,\sigma').
\end{equation}

During the time-evolution of the MPEMs (Sec.~\ref{sec:mpemAlgo}) it is natural to normalize the edge messages according to the 2-norm. In that case, it can be necessary to add a renormalization in Eq.~\eqref{eq:mpem_observe} in order to avoid the generation of very large matrix elements, e.g., by replacing in each step $E^{(s)}$ by $E^{(s)}/\norm{E^{(s)}}$.

\section{An improved truncation scheme}\label{sec:dmTrunc}
In order to recast the evolved state \eqref{eq:MPEM_C} into canonical form \eqref{eq:MPEM_B} and truncate bond dimensions, the first step in Sec.~\ref{sec:B_from_C} was to reorthonormalize the tensors ($C\to\tilde{C}$) in a sweep from right to left before doing truncations in a subsequent sweep from left to right ($\tilde{C}\to\bar{C}$). The preparatory first sweep can be avoided as described in the following. This also reduces computation costs substantially from $\mc{O}\big(M^{3z-3}\big)$ to $\mc{O}(M^{2z-1})$, where $M$ denotes MPEM bond dimensions and $z$ denotes vertex degrees.

In extension of Sec.~\ref{sec:MPS_trunc}, let us first discuss how controlled truncations (of Schmidt coefficients) can be done when the basis $\{|a\ket\}$ of the left part $\mc{L}$, containing sites $[1,\ell]$, is orthonormal while the basis $\{|b\ket\}$ for the right part $\mc{R}$, containing sites $[\ell+1,L]$, is \emph{not} orthonormal. Recall that the purpose of the preparatory first sweep in Sec.~\ref{sec:B_from_C} was to orthonormalize \emph{both} bases. For notational simplicity, let us again use the example of a quantum many-body system with $L$ lattice sites. We are given a state
\begin{equation}\label{eq:psiBeforeTrunc2}
	|\psi\ket=:\sum_{a,b}\psi_{a,b}|a\ket\otimes|b\ket
\end{equation}
with orthonormal states $\bra a|a'\ket=\delta_{a,a'}$ for the left part and arbitrary states $|b\ket$ for the right part.
Now one can find a good approximation $|\psi_\trunc\ket$ of $|\psi\ket$ (with respect to the 2-norm distance) by diagonalization of the reduced density matrix $\hat{\varrho}_\mc{L}$ for the left part. This density matrix,
\begin{equation}
	\hat{\varrho}_\mc{L}\equiv\Tr_\mc{R}|\psi\ket\bra\psi|=\sum_{a,a',b,b'}\psi_{a,b}|a\ket\bra b'|b\ket\bra a'| \psi^*_{a',b'},
\end{equation}
is obtained by a partial trace over $\mc{R}$. With the overlap matrix $F_{b,b'}:=\bra b'|b\ket$ we have $[\varrho_\mc{L}]_{a,a'}:=\bra a|\hat{\varrho}_\mc{L}|a'\ket=[\psi F\psi^\dag]_{a,a'}$ which can be diagonalized according to
\begin{equation}\label{eq:psiTrunc2a}
	\varrho_\mc{L}=\psi F\psi^\dag=:U\Lambda^2 U^\dag,
\end{equation}
where, as in Sec.~\ref{sec:MPS_trunc}, $U$ is a unitary matrix and $\Lambda$ is the diagonal matrix containing descendingly ordered Schmidt coefficients $\lambda_1\geq\lambda_2\geq\dots\geq \lambda_{M'}$ (square roots of density matrix eigenvalues).
\begin{subequations}
\begin{equation}
	|\psi\ket=\sum_{a,b,k}U_{a,k} R_{k,b}|a\ket\otimes|b\ket
	 =\sum_{k=1}^{M'} |k_\mc{L}\ket\otimes|\tilde{k}_\mc{R}\ket,
\end{equation}
where $\{|k_\mc{L}\ket:=\sum_a U_{a,k}|a\ket\}$ is an orthonormal basis for $\mc{L}$ and $\{|\tilde{k}_\mc{R}\ket:=\sum_b R_{k,b}|b\ket\}$ are orthogonal states with norms $\lambda_k$ for $\mc{R}$ with $R:=U^\dag \psi$.
\begin{figure}[t]
\label{fig:overlapMatrix}
\includegraphics[width=1\textwidth]{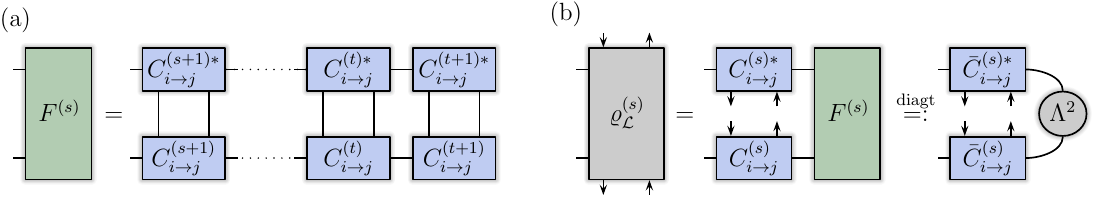}
\caption{(a) Overlap matrices $F^{(s)}$ are needed in the density matrix truncation scheme discussed in Sec.~\ref{sec:dmTrunc}. It allows to truncate without a preparatory orthonormalization sweep and substantially reduces computation costs. Overlap matrices can be computed using the iteration \eqref{eq:overlapMatrices}. (b) From $F^{(s)}$ and tensor $C^{(s)}$, one obtains the reduced density matrix $\varrho^{(s)}$ [Eq.~\eqref{eq:rhoL}] and its diagonalization and truncation \eqref{eq:rhoLdiagt} yields tensor $\bar{C}^{(s)}$ that obeys the left orthonormality constraint and the singular values $\Lambda^{(s)}$ employed in the truncation.}
\end{figure}
The approximation after truncating all Schmidt coefficients $\lambda_k$ with $k>M$ is
\begin{equation}\label{eq:psiTrunc2}
	|\psi_\trunc\ket=\sum_{k=1}^{M} |k_\mc{L}\ket\otimes|\tilde{k}_\mc{R}\ket
	\quad \text{with error} \ \
	\norm{\psi-\psi_\trunc}^2=\!\!\sum_{k=M+1}^{M'}\|\tilde{k}_\mc{R}\|^2=\!\!\sum_{k=M+1}^{M'}\lambda_k^2,
\end{equation}
\end{subequations}
The orthogonality and norms of the states $\{|\tilde{k}_\mc{R}\ket\}$ follow as
\begin{equation}
	\bra\tilde{k}_\mc{R}'|\tilde{k}_\mc{R}\ket
	 =\sum_{b,b'}R^*_{k',b'}\bra b'|b\ket R_{k,b}
	 =[RFR^\dag]_{k,k'}
	 =[U^\dag\psi F \psi^\dag U]_{k,k'}
	 =[\Lambda^2]_{k,k'}=\delta_{k,k'}\lambda_k^2.
\end{equation}
The truncation \eqref{eq:psiTrunc2} is suitable for an MPS $|\psi\ket$ (and analogously for MPEM) given in the form 
\begin{equation}\label{eq:ONconstr2}
	|\psi\ket = \sum_{\vec{n}} Y_1^{n_1}\dotsb Y_\ell^{n_\ell}\cdot A_{\ell+1}^{n_{\ell+1}}\dotsb A_L^{n_L}\,|\vec{n}\ket \quad\text{with}\quad \sum_n (Y_i^n)^\dag Y_i^n = \id.
\end{equation}
This is the form \eqref{eq:psiBeforeTrunc2} with orthonormal basis states $|a\ket=\sum_{n_1,\dotsc,n_\ell} \big[Y_1^{n_1}\dotsb Y_\ell^{n_\ell}\big]_{1,a}\,|n_1\dots n_\ell\ket$ for $\mc{L}$ [cf.\ Eq.~\eqref{eq:ONleft}], non-orthogonal states $|b\ket=\sum_{n_{\ell+1},\dotsc,n_L} \big[A_{\ell+1}^{n_{\ell+1}}\dotsb A_L^{n_L}\big]_{b,1}\,|n_{\ell+1}\dots n_L\ket$ for $\mc{R}$, and $\psi=\id$. The overlap matrix $F$ can be computed in an iteration with $F_L:=\id$ and $F_j:=\sum_n A^n_{j+1} F_{j+1} (A^n_{j+1})^\dag$ for $j=L-1,\dotsc,\ell$. With a diagonalization of $F\equiv F_{\ell}$, as in Eq.~\eqref{eq:psiTrunc2a}, we then obtain an optimally truncated state $|\psi_\trunc\ket$ in MPS form.

With this type of truncation, the preparatory first orthonormalization sweep for the evolved MPEM (right to left, $C\to\tilde{C}$), described in Sec.~\ref{sec:B_from_C}, can be avoided and computation costs can be reduced substantially. Going from right to left, one can first compute all overlap matrices $F^{(s)}$ for the evolved MPEM \eqref{eq:MPEM_C} with
\begin{equation}\label{eq:overlapMatrices}
	F^{(t+1)}:=\id\quad \text{and} \quad
	F^{(s)}:=\sum_{\sigma_i,\sigma_j} C^{(s+1)}_{i\to j}(\sigma_i|\sigma_j) F^{(s+1)} \big(C^{(s+1)}_{i\to j}(\sigma_i|\sigma_j)\big)^\dag
	\quad\text{for}\quad
	s=t,\dotsc,0
\end{equation}
as illustrated in Fig.~\ref{fig:overlapMatrix}(a).
With these, we can directly truncate the evolved MPEM, in a sweep from left to right. At the left boundary, we start with
\begin{subequations}\label{eq:dmTruncL}
\begin{gather}
	\varrho^{(0)}_\mc{L}(\sigma_i,\sigma_i') = C^{(0)}_{i\to j}(\sigma_i) F^{(0)} \big(C^{(0)}_{i\to j}(\sigma_i')\big)^\dag
	\stackrel{\diagt}{=:} \bar{C}^{(0)}_{i\to j}(\sigma_i) \big(\Lambda^{(0)}\big)^2 \big(\bar{C}^{(0)}_{i\to j}(\sigma_i')\big)^\dag\\
	\text{and}\quad
	X^{(0)}:=\sum_{\sigma_i} \big(\bar{C}^{(0)}_{i\to j}(\sigma_i)\big)^\dag C^{(0)}_{i\to j}(\sigma_i),
\end{gather}
\end{subequations}
where we diagonalize and truncate ($\diagt$) with the diagonalization as in Eq.~\eqref{eq:psiTrunc2a} and the truncation as in Eq.~\eqref{eq:psiTrunc2}. The new tensor $\bar{C}^{(0)}_{i\to j}$ obeys the left orthonormality constraint [cf.\ Eq.~\eqref{eq:ONconstr2}] and we need to include the matrix $X^{(0)}$ right of $\bar{C}^{(0)}_{i\to j}$ to keep the matrix product invariant such that
$C^{(0)}_{i\to j}(\sigma_i^{0}) \stackrel{\trunc}{\approx} \bar{C}^{(0)}_{i\to j}(\sigma_i^{0}) X^{(0)}$ in Eq.~\eqref{eq:MPEM_C}. For the bulk tensors $s=1,2,\dotsc,t$, we continue with
\begin{subequations}\label{eq:dmTruncBulk}
\begin{gather}\label{eq:rhoL}
	\varrho^{(s)}_\mc{L}(\sigma_i\sigma_j,\sigma_i'\sigma_j') = X^{(s-1)} C^{(s)}_{i\to j}(\sigma_i|\sigma_j) F^{(s)} \big(X^{(s-1)}C^{(s)}_{i\to j}(\sigma_i'|\sigma_j')\big)^\dag\\\label{eq:rhoLdiagt}
	\phantom{aaaaaaaaaa}\,\,\,
	\stackrel{\diagt}{=:} \bar{C}^{(s)}_{i\to j}(\sigma_i|\sigma_j) \big(\Lambda^{(s)}\big)^2 \big(\bar{C}^{(s)}_{i\to j}(\sigma_i'|\sigma_j')\big)^\dag\\
	\text{and}\quad
	X^{(s)}:=\sum_{\sigma_i,\sigma_j} \big(\bar{C}^{(s)}_{i\to j}(\sigma_i|\sigma_j)\big)^\dag C^{(s)}_{i\to j}(\sigma_i|\sigma_j).
\end{gather}
\end{subequations}
See Fig.~\ref{fig:overlapMatrix}(b).
We end at the right boundary with $\bar{C}^{(t+1)}_{i\to j}(\sigma_i|\sigma_j):=X^{(t)}C^{(t+1)}_{i\to j}(\sigma_i|\sigma_j)$ such that all $\bar{C}$-tensors except $\bar{C}^{(t+1)}$ now obey the left orthonormality constraint.

Truncating in this way, we arrive at Eq.~\eqref{eq:MPEM_Ctrunc} of the conceptually simpler truncation scheme of Sec.~\ref{sec:B_from_C} and can continue in the same way as described there, rearranging vertex variables in the evolved MPEM to bring it into the canonical form \eqref{eq:MPEM_B}. The major advantage of the more elaborate density matrix truncation scheme is the reduced computation cost. With bond dimensions $M$ of the MPEM \eqref{eq:MPEM} before the evolution step, the $C$-tensors of the exactly evolved MPEM \eqref{eq:MPEM_C} have increased bond dimensions $M^{z-1}$, where $z$ is the degree of vertex $i$. The computationally most expensive step in the simple truncation scheme of Sec.~\ref{sec:B_from_C} is the singular value decomposition \eqref{eq:Cortho} for the orthonormalization of the $C$-tensors; the cost scales as $\mc{O}\big(M^{3z-3}\big)$. The most expensive steps in the more efficient density matrix truncation scheme are the computations of overlap matrices \eqref{eq:overlapMatrices} and the reduced density matrices \eqref{eq:rhoL}. Exploiting the structure of the $C$-tensors \eqref{eq:C}, these operations can be accomplished with a cost $\mc{O}(M^{2z-1})$ if we assume that the bond dimensions of tensors $\bar{C}$ of the evolved MPEM after truncation are similar to that of the $A$-tensors in the original MPEM \eqref{eq:MPEM}, i.e., approximately $M$. This is generally the case. When we fix a truncation threshold $\varepsilon$ for the discarded 2-norm weight as discussed in Sec.~\ref{sec:B_from_C}, bond dimensions evolve smoothly in time.

\section{Computation costs in Glauber-Ising dynamics}\label{sec:example}
As the number of possible trajectories on a vertex increases exponentially in time, the computation costs for exact dynamic belief propagation \eqref{eq:dynamicCavityEq} also grow exponentially. As discussed in the introduction, this has, so far, substantially limited the applicability of the dynamic cavity method. The MPEM approach allows for a controlled reduction of the computational complexity, exploiting the decay of temporal correlations to truncate unimportant components of the edge messages.
\begin{figure}[t]
\label{fig:GlauberIsing}
\includegraphics[width=0.8\textwidth]{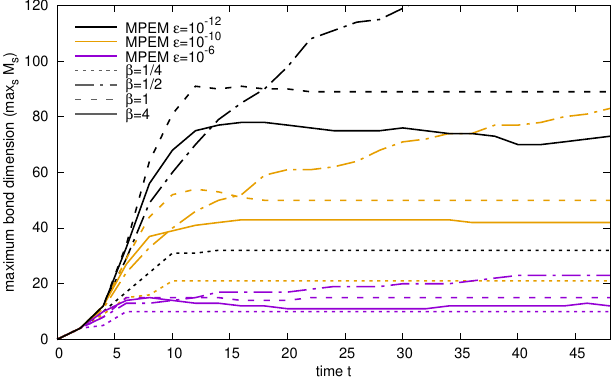}
\caption{Evolution of maximum MPEM bond dimensions $\max_s M_s$ for Glauber dynamics of the kinetic Ising model \eqref{eq:GlauberIsing} on $z=3$ random regular graphs. Truncation thresholds, indicated by color, were fixed to $\varepsilon=10^{-6},10^{-10}$, and $10^{-12}$, respectively. Bond dimensions are shown for inverse temperatures $\beta=4,1,1/2$, and $1/4$ (indicated by dash type). Bond dimensions increase with decreasing truncation threshold. They become constant at longer times which implies that computation costs only grow linearly in time.}
\end{figure}

Concerning computation costs, a decisive question is now, what bond dimensions $M_s$ in MPEMs are required to achieve a certain approximation accuracy in comparison to the exact evolution. Also, how does the required bond dimension evolve with time? Generally, it is to be expected that the maximum bond dimension, $\max_s M_s$, will converge to a constant as a function of time if connected temporal correlations decay exponentially. In simulations, it is often favorable not to fix the truncation dimension, but to fix instead a threshold $\varepsilon$ and to truncate all singular values $\lambda_k$ with $\lambda_k^2 <\varepsilon \cdot\sum_q\lambda_q^2$. The threshold $\varepsilon$ controls the 2-norm loss in each truncation. Allowing bond dimensions to evolve accordingly, avoids wasting computation time and gives a measure for the information-theoretic complexity of the edge messages.

As an example, we consider Glauber dynamics of the kinetic Ising model \cite{Glauber1963-4} on random regular graphs with vertex degree $z=3$ in the thermodynamic limit. Specifically, Ising spins interact ferromagnetically with local transition matrices
\begin{equation}\label{eq:GlauberIsing}
	w_i(\sigma_i^{t+1}|\vs_{\partial i}^{t})=\frac{1}{Z}\exp\big(\beta\sum_{j\in\partial i} \sigma_i^{t+1}\sigma_j^{t}\big).
\end{equation}
At time $t=0$, all spins have magnetization $\bra\sigma^0_i\ket=1/2$, i.e., $p_i(\uparrow{}\!)=3/4$.

Figure~\ref{fig:GlauberIsing} shows the evolution of maximum bond dimensions with time. They increase with decreasing truncation threshold $\varepsilon$. As expected, they become constant at longer times. The number of tensors $A^{(0)},\dotsc,A^{(t+1)}$ in an MPEM increases linearly with $t$, and we need to sweep a few times through the matrix product in each iteration. With converging bond dimensions, this implies that the computation cost per iteration grow only linearly in time, instead of exponentially. For fixed $\varepsilon$, the largest bond dimensions are needed for inverse temperature $\beta=1/2$. This is so, because it is the $\beta$ closest to the phase transition in the Ising model. For this temperature, the bond dimensions do actually not yet show saturation on the time interval displayed in Fig.~\ref{fig:GlauberIsing}. They will converge at larger $t$.

\section{Models with vertex-state dependence}\label{sec:continuum}
Except for the algorithms described in Secs.~\ref{sec:mpemAlgo} and \ref{sec:dmTrunc} our description was generic in the sense that the local transition matrices $w_i=w_i(\sigma_i^{t+1}|\vs_{\partial i}^{t})$ were allowed to depend on the time-$t$ state $\sigma_i^{t}$ of vertex $i$ itself, in addition to the states on neighboring vertices. However, in Secs.~\ref{sec:mpemAlgo} and \ref{sec:dmTrunc} we considered the case where $\partial i$ does not contain $i$ itself, i.e., that $w_i$ is independent of $\sigma_i^{t}$. Here, we generalize to MPEM algorithms that allow for the dependence on $\sigma_i^{t}$. This is important for many applications like stochastic models for the dynamics of infectious diseases \cite{Murray1989,Dangerfield2009-6,Karrer2010-82} or of opinions is social networks \cite{Castellano2009-81}. Naturally, transition matrices with dependence on $\sigma_i^{t}$ also arise in time-discretized versions of stochastic continuous-time dynamics, i.e., all models that have a well-defined continuum-time limit as, for example, the Glauber dynamics of Ising spin systems \cite{Glauber1963-4}. In particular, when decreasing the time step $\Delta t$, one should have
\begin{equation}
	w_i(\sigma_i^{t+1}|\vs_{\partial i}^{t})\to\delta_{\sigma_i^{t+1},\sigma_i^{t}}
	\quad\text{for}\quad
	\Delta t\to 0.
\end{equation}

Vertex $i$ is now contained in its neighborhood $\partial i=\{i,j,k_1,\dotsc,k_{z-1}\}$, where $z$ denotes the vertex degree. In this more general scenario, the evolved MPEM \eqref{eq:MPEM_C} cannot be constructed in an entirely local fashion anymore. Instead, the $\tilde{C}$-tensors are constructed in a sweep from right ($s=t+1$) to left ($s=0$), imposing at the same time the right orthonormality constraint, $\sum_{\sigma,\sigma'}\tilde{C}^{(s)}_{i\to j}(\sigma|\sigma')\big[\tilde{C}^{(s)}_{i\to j}(\sigma|\sigma')\big]^\dag=\id$ [cf.\ Eq.~\eqref{eq:ONconstr}].

We start at the right boundary ($s=t+1$) by doing a singular value decomposition (SVD) of the tensor
\begin{equation*}
	C^{(t+1)}_{i\to j}(\sigma_i^{t+1}|\sigma_i^{t},\sigma_j^{t})
	:=\!\sum_{\sigma_{k_1}^{t}\dotsc\sigma_{k_{z-1}}^{t}}\!
w(\sigma_i^{t+1}|\vs_{\partial i}^{t}) \Big[\bigotimes_{k=1}^{z-1} A^{(t+1)}_{k_n\to i}(\sigma_{k_n}^{t})\Big]
	\stackrel{\SVD}{=:} U^{(t+1)}(\sigma_i^{t})\Lambda^{(t+1)}\tilde{C}^{(t+1)}_{i\to j}(\sigma_i^{t+1}|\sigma_j^{t})
\end{equation*}
in order to obtain the tensor $\tilde{C}^{(t+1)}(\sigma_i^{t+1}|\sigma_j^{t})$, where $\Lambda^{(t+1)}$ is the diagonal matrix of singular values, $\tilde{C}^{(t+1)}_{i\to j}$ obeys the right orthonormality constraint, and, similarly, $\sum_{\sigma}\big[U^{(t+1)}(\sigma)\big]^\dag U^{(t+1)}(\sigma)=\id$. For time $s=t$, the process continues with
\begin{align}\nonumber
	C^{(t)}_{i\to j}(\sigma_i^{t}|\sigma_i^{t-1},\sigma_j^{t-1})
	&:=\sum_{\sigma_{k_1}^{t-1}\dotsc\sigma_{k_{z-1}}^{t-1}}
	    w(\sigma_i^{t}|\vs_{\partial i}^{t-1})
	  \Big[\bigotimes_{n=1}^{z-1} A^{(t)}_{k_n\to i}(\sigma_{k_n}^{t-1})\Big]U^{(t+1)}(\sigma_i^{t})\Lambda^{(t+1)}\\
	  \label{eq:continuumR}
	&\stackrel{\SVD}{=:}U^{(t)}(\sigma_i^{t-1})\Lambda^{(t)}\tilde{C}^{(t)}_{i\to j}(\sigma_i^{t}|\sigma_j^{t-1}).
\end{align}
For the bulk tensors, $s=t,t-1,\dotsc,1$, the corresponding equations read very similarly
\begin{align}\nonumber
	C^{(s)}_{i\to j}(\sigma_i^{s}|\sigma_i^{s-1},\sigma_j^{s-1})
	&:=\sum_{\sigma_{k_1}^{s-1}\dotsc\sigma_{k_{z-1}}^{s-1}}
	    w(\sigma_i^{s}|\vs_{\partial i}^{s-1})
	  \Big[\bigotimes_{n=1}^{z-1} A^{(s)}_{k_n\to i}(\sigma_{k_n}^{s-1}|\sigma_{i}^{s})\Big]U^{(s+1)}(\sigma_i^{s})\Lambda^{(s+1)}\\
	  \label{eq:continuumBulk}
	&\stackrel{\SVD}{=:}U^{(s)}(\sigma_i^{s-1})\Lambda^{(s)}\tilde{C}^{(s)}_{i\to j}(\sigma_i^{s}|\sigma_j^{s-1})
\end{align}
\begin{figure}[b]
\label{fig:MPEM_orthonormalizeCont}
\includegraphics[width=0.9\textwidth]{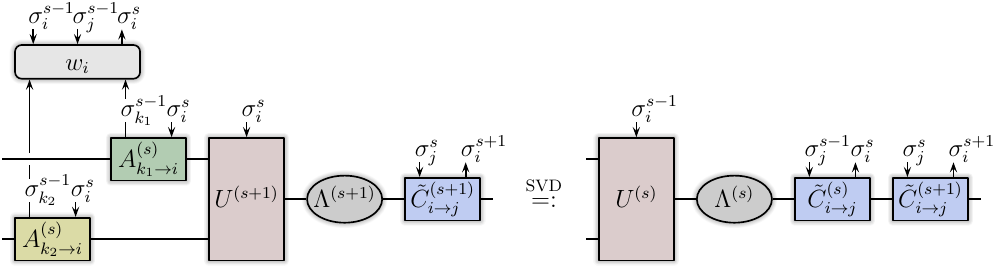}
\caption{In Sec.~\ref{sec:continuum}, we discuss the case where the transition matrix $w_i=w_i(\sigma_i^{t+1}|\vs_{\partial i}^{t})$ for vertex $i$ is allowed to also depend on the time-$t$ state $\sigma_i^{t}$ of vertex $i$ itself, instead of just depending on states of nearest neighbors. In this case, $C$-tensors of the evolved MPEM \eqref{eq:MPEM_C} cannot be constructed in an entirely local fashion anymore. Instead, the $\tilde{C}$-tensors that also obey the right orthonormality constraint are constructed in a sweep from right ($s=t+1$) to left ($s=0$), sequentially doing singular value decompositions and contracting with $A$-tensors from MPEMs of the previous time step. The iteration, corresponding to Eq.~\eqref{eq:continuumBulk}, is shown for a vertex with degree $z=3$. Compare to Fig.~\ref{fig:MPEM_orthonormalize} for the case without vertex-state dependence.}
\end{figure}
as illustrated in Fig.~\ref{fig:MPEM_orthonormalizeCont}. The process ends at the left boundary with the assignment
\begin{equation}\label{eq:continuumL}
	\tilde{C}^{(0)}_{i\to j}(\sigma_i^{0}):=
	    p_i(\sigma_i^{0})
	  \Big[\bigotimes_{n=1}^{z-1} A^{(0)}_{k_n\to i}(\sigma_{i}^{0})\Big]U^{(1)}(\sigma_i^{0})\Lambda^{(1)}.
\end{equation}
The remaining truncation and vertex variable reordering sweeps, done in order to transform the non-canonical form \eqref{eq:MPEM_C} into the canonical form \eqref{eq:MPEM_B}, can be executed exactly as discussed in the somewhat simpler situation covered in Sec.~\ref{sec:B_from_C}.

The more efficient density matrix truncation scheme discussed in Sec.~\ref{sec:dmTrunc}, similarly, can be adapted to the situation where transition matrix $w_i=w_i(\sigma_i^{t+1}|\vs_{\partial i}^{t})$ depends on $\sigma_i^{t}$. Like the unitaries $U^{(s)}$ in Eqs.~\eqref{eq:continuumR}-\eqref{eq:continuumL}, the $X^{(s)}$-matrices in Eqs.~\eqref{eq:dmTruncL} and \eqref{eq:dmTruncBulk} of the density matrix truncation scheme will then depend on $\sigma_i^{s-1}$.

Continuous-time stochastic dynamics on locally tree-like networks can be simulated with the described approach after discretization of the time axis using a small time step $\Delta t$. It should also be possible to, alternatively, work with continuous matrix products \cite{Verstraete2010-104,Osborne2010-105} as introduced in the context of condensed matter physics . This however, entails some technical difficulties and one can expect that using discrete-time MPEMs and a small time step should be the best option. Similarly, quantum many-body systems in continuous real-space have been treated efficiently using MPS with a sufficiently fine space discretization \cite{Stoudenmire2012-109,Dolfi2012-109}. Other approximative schemes to simulate continuous-time stochastic dynamics are the dynamical replica analysis \cite{Mozeika2008-41} that captures macroscopic observables and the cavity master equation method \cite{Aurell2017-95} that operates on local conditional probabilities.

\section{Discussion}
The described MPEM algorithm for the simulation of stochastic dynamics is based on the cavity method, applicable for locally tree-like interaction graphs, and on the matrix product approximation for edge messages which exploits the decay of (connected) temporal correlations.
The MPEM method lifts restrictions of earlier approaches for the solution of dynamic cavity equations, mentioned in the introduction. It can also be used to simulate in the thermodynamic limit. Compared to Monte Carlo simulations, errors decrease much faster as a function of the 
invested computation time. This has been demonstrated for Glauber dynamics of the kinetic Ising model in Ref.~\cite{Barthel2018-97}. Hence, the MPEM method is particularly suited for the precise analysis of temporal correlations, decay processes, and low-probability events. For the Glauber-Ising dynamics and fixed truncation thresholds, we have seen here that MPEM bond dimensions converge to a constant as a function of time. This should generally apply when connected temporal correlations decay exponentially. It implies that required computation costs per iteration grow only linearly instead of exponentially in time.

There are several ways in which the MPEM method can be developed further. Here, we have discussed a more efficient truncation scheme that reduces computation costs considerably from $\mc{O}(M^{3z-3})$ to $\mc{O}(M^{2z-1})$, and we have generalized the method to models where the transition probabilities for a vertex depend explicitly on the vertex state at the previous time step, in addition to the states for its nearest neighbors.

Efficient codes for the MPEM algorithms, including the optimized truncation scheme of Sec.~\ref{sec:dmTrunc}, are available from the author and under \href{http://www.manyparticle.org/~barthel/mpem}{www.manyparticle.org/$\sim$barthel/mpem}. They were, for example, employed for the simulations in Ref.~\cite{Barthel2018-97}. We are also happy to collaborate on specific applications.

I gratefully acknowledge discussions with Silvio Franz and Caterina De~Bacco that initiated this work and support through US Department of Energy grant DE-SC0019449.

\end{document}